\documentclass[a4paper,pra,showpacs,twocolumn]{revtex4}
\usepackage{amsfonts}
\usepackage{amssymb}
\usepackage{amsmath}
\usepackage{calc}
\usepackage{graphicx}
\begin{document}
\title{Adiabatic creation of coherent superposition states via multiple
  intermediate states}

\author{A. Karpati and Z. Kis}

\affiliation{
  Department of Nonlinear and Quantum Optics, \\
  Research Institute for Solid State Physics and Optics, \\ 
  Hungarian Academy of Sciences, P.O. Box 49, H-1525 Budapest, Hungary\\
  }

\date{\today}

\begin{abstract}
  We consider an adiabatic population transfer process that resembles the well
  established stimulated Raman adiabatic passage (STIRAP). In our system, the
  states have nonzero angular momentums $J$, therefore, the coupling laser
  fields induce transitions among the magnetic sublevels of the states. In
  particular, we discuss the possibility of creating coherent superposition
  states in a system with coupling pattern $J=0\Leftrightarrow J=1$ and
  $J=1\Leftrightarrow J=2$. Initially, the system is in the $J=0$ state. We
  show that by two delayed, overlapping laser pulses it is possible to create
  any final superposition state of the magnetic sublevels $|2,-2\rangle$,
  $|2,0\rangle$, $|2,+2\rangle$.  Moreover, we find that the relative phases
  of the applied pulses influence not only the phases of the final
  superposition state but the probability amplitudes as well. We show that if
  we fix the shape and the time-delay between the pulses, the final state
  space can be entirely covered by varying the polarizations and relative
  phases of the two pulses.  Performing numerical simulations we find that our
  transfer process is nearly adiabatic for the whole parameter set.
\end{abstract}


\maketitle

\section{Introduction}

Adiabatic methods in quantum mechanics play a fundamental role in the
treatment of level crossing problems. The famous Landau-Zener system \cite{LZ}
has been followed by several other models which were defined in two-level
systems \cite{Rosen,Demkov,Hioe,Suominen}. These models induced many
applications in atomic and molecular physics, for recent review see Ref.
\cite{vita}.

It is common in all of the adiabatic systems that the Hamiltonian is
time-dependent. The instantaneous eigenstates define an adiabatic basis.
Suppose that at the initial time, one of the adiabatic states coincides with
the initial state of the system. In most cases the adiabatic state that
belongs to the eigenvalue zero is utilized for the transfer. If the system
varies with time slowly enough, then its state vector follows adiabatically
the adiabatic eigenstate and it will go over smoothly to the desired final
state.  The adiabaticity of the process guarantees its robustness with respect
to fluctuations in the parameters of the couplings. In the limit of ideal
adiabatic transfer, the other eigenstates are not invoked in the evolution and
their population remains zero throughout the whole time.  The non-adiabatic
corrections tend to involve these states, but the magnitudes of such couplings
are supposed to be small compared with their energy separation from the
adiabatic states.

During the last decade, a promising new population transfer technique was
introduced: the stimulated Raman adiabatic passage (STIRAP) process realizes
efficient population transfer in three-level systems, see Ref.  \cite{vitb}
and references therein. Moreover, it turned out that this scheme can be
applied successfully to the problem of manipulating and creating coherent
state superpositions. Such superpositions are the desired initial states for
many modern quantum applications including information processing and
communication. The original STIRAP process has thus been utilized to create
coherent superpositions in three- and four-level systems
\cite{marte,una1,theuer,una2,Chang} and to prepare $N$-component maximally
coherent superposition states \cite{una3}. The four-level (tripod) STIRAP
establishes such applications as qubit rotation \cite{kisc} and density-matrix
measurement \cite{kisd}.

Recently, a multilevel STIRAP scheme has been proposed \cite{kisa, kisb} to
create $N-$level coherent superposition states. This model is particularly
interesting, since the Hamiltonian has several adiabatic eigenstates which
belong to the eigenvalue zero. The diabatic couplings among these eigenstates
influence significantly the dynamics of the system. However, the eigenstates
in the zero eigenvalue subspace are decoupled from the eigenstates belonging
to nonzero eigenvalues provided the time-evolution is adiabatic. In Ref.
\cite{kisa} the Optimal Control Theory has been applied to find the optimal
pulse sequences which create a prescribed final superposition state. It has
also been verified that the transfer process occurs in the zero-eigenvalue
subspace. In a subsequent work \cite{kisb} an analytic solution has been
presented for a five-level system.

When the STIRAP concept is realized in atoms or molecules one encounters that
the quantum states with angular momentum $J$ have magnetic sublevels and these
sublevels are coupled by the applied laser pulses, and a rather complicated
coupling-pattern may occur. This situation has been studied in Refs.
\cite{Shore,Martin1,Martin2}. In these papers the authors have discussed the
possibility of transferring population from a well defined initial state to a
single but arbitrarily chosen magnetic sublevel. 

An other realistic problem is the presence of multiple or continuum
intermediate states \cite{Carroll,Nakajima,vitc,Yatsenko, vitd, Paspalakis,
  una4, una5}. In Ref. \cite{Carroll} the continuum has been modeled by
infinite number of equidistant discrete levels which were coupled by unique
couplings to a single initial and final state.  Later, it has been shown that
the previous model is too simple, the physics is more complicated in a
realistic situation \cite{Nakajima}. A detailed study of the discrete model
has been presented in Ref. \cite{vitc}. It has also been shown that efficient
population transfer is indeed possible through a continuum \cite{Yatsenko,
  vitd, Paspalakis, una4, una5}.

In this paper we will study a multilevel $\Lambda$ scheme similar to the one
discussed in Refs. \cite{Shore,Martin1,Martin2}. However, in contrast with the
previous works where the population transfer took place between a single
initial and a single final state, our aim is to create a multilevel final
superposition state. In our model, the initial state has angular momentum
$J\smash{=}0$, the intermediate state has $J\smash{=}1$ and the final
superposition state is formed from the magnetic sublevels of a state
$J\smash{=}2$. We will design a STIRAP-like process, i.e. the population
transfer is realized by delayed, overlapping laser pulses.  We define the
Hamiltonian corresponding to the process, separate the adiabatic states
belonging to the eigenvalue zero and find the conditions for the successful
population transfer. We will study how does the final state depend on the
relative phases of the applied laser pulses. The robustness of the transfer
process will also be discussed.

The organization of the paper is as follows: In Section \ref{model} the
physical model of our six-level STIRAP is presented and the corresponding
Hamiltonian is introduced.  The Hilbert space is separated to a
zero-eigenvalue subspace and to a nonzero-eigenvalue subspace. The eigenstates
belonging to the zero eigenvalue are determined explicitly.  In Section
\ref{poptrans} we study the possibility of population transfer by means of
delayed laser pulses. We show numerically that nearly adiabatic evolution is
possible. We also discuss the robustness of the transfer process. We summarize
the results in Section \ref{sum}.

\section{Six-level system}\label{model}
\begin{figure}
\includegraphics[width=\textwidth/2-1cm]{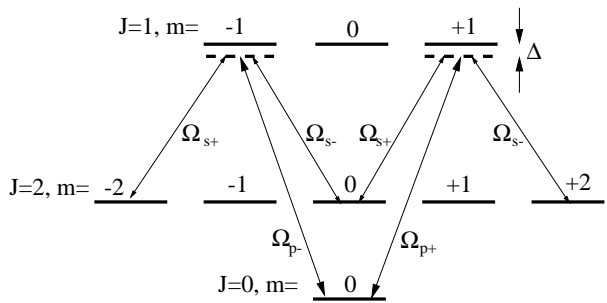}
\caption{The six-level $\Lambda$ system. The levels $|2,\pm1\rangle$
  and $|1,0\rangle$ are not coupled to other levels. The remaining levels form
  the six-level $\Lambda$ configuration. Levels with $J\smash{=} 0$ and
  $J\smash{=} 2$ are coupled through the levels with $J\smash{=}1$ by the pump
  (p) and Stokes (S) laser pulses with right- and left-hand circular
  polarization $(+/-)$.  The corresponding Rabi frequencies are
  $\Omega_{S\pm}$ and $\Omega_{p\pm}$, respectively. The pulses are at
  two-photon resonance, but may be detuned from the $J\smash{=}1$ levels by a
  certain detuning $\Delta$.  Only the level $|0,0\rangle$ is populated
  initially.  }
\label{fig:levels}
\end{figure}

Let us consider the six-level $\Lambda$ configuration shown in Fig.
\ref{fig:levels}.  The system consists of the magnetic sublevels of states
with total angular momentum $J\smash{=}0$, $J\smash{=}1$ and $J\smash{=}2$.
The $J\smash{=}1$ states have higher energy than the $J\smash{=}0$ and
$J\smash{=}2$ states. Initially, only the $J\smash{=}0$ state is populated.
Our goal is to transfer this population to the sublevels of the $J\smash{=}2$
state and to obtain a prescribed final superposition state. We are going to
design a STIRAP-like population transfer process. To this end, we apply two
sets of laser pulses, the corresponding Rabi frequencies are denoted by
$\Omega_{S\pm}$ and $\Omega_{p\pm}$. The subscripts $S$ and $p$ refer to
``Stokes'' and ``pulse'', while the indices $+/-$ correspond to right- and
left-hand circular polarizations, respectively.  Obviously, in this
configuration only the sublevels with $m\smash{=}-2, 0, +2$ can be populated
in the final state. This linkage can be realized experimentally in Neon atoms,
see Ref.~\cite{Martin2}. The more general case, where all magnetic sublevels
of the $J\smash{=}2$ state are included into the final superposition state
will be discussed elsewhere.  The frequency and the phase of the laser pulses
are kept constant during the interaction, only their amplitudes are
time-dependent.

In our model we have to determine the Rabi frequencies in an angular-momentum
basis \cite{bruce}. According to the Wigner-Eckart theorem, the general
formula for the Rabi frequency between two levels of angular momenta $J_1$ and
$J_2$ and magnetic quantum numbers $M_1$ and $M_2$ is given by
\begin{eqnarray}\label{Rabi}
  \hbar\Omega(J_1 M_1; J_2 M_2)\!&=&\!-{\cal E}^*(-1)^{J_1-M_1}
  (J_1||\hat{d}||J_2) \nonumber\\ 
  &\times&\!\!\sum_{s=-1}^{+1} (-1)^s \epsilon_{-s}\left[
    \begin{array}{ccc}
      J_1 & 1 & J_2 \\
      -M_1 & s & M_2 
    \end{array}
  \right]\,,
\end{eqnarray}
where the electric field strength is ${\cal E}$, the reduced dipole matrix
element is given by $(J_1||\hat{d}||J_2)$, $\epsilon_s$ are the irreducible
tensor operator components of the polarization vector of the field, and
$[\dots]$ denotes the $3-j$ symbol. We apply left- and right-hand circularly
polarized light, therefore, in the summation $s=\pm1$ and $\epsilon_{\pm1}=1$.

The two pump pulses couple the level $|00\rangle$ to the levels
$|1,\pm1\rangle$. The corresponding Rabi frequencies can be expressed as
$\Omega_{p\pm}= {\rm const}_1\times {\cal E}_{p\pm}^*$.  The right-hand
circularly polarized Stokes pulse couple the levels
$|1,-1\rangle\Leftrightarrow |2,-2\rangle$ and $|1,+1\rangle\Leftrightarrow
|2,0\rangle $. For the Rabi frequencies we obtain
\begin{subequations}
\begin{eqnarray}
  \Omega_{S+}(1, -1; 2, -2)&=& q\times {\rm const}_2\times {\cal E}_{S+}^*
  \,,\\ 
  \Omega_{S+}(1, +1; 2, 0)&=& {\rm const}_2\times {\cal E}_{S+}^* \,,
\end{eqnarray}
\end{subequations}
where $q$ is the ratio of the corresponding $3-j$ symbols. In our case
$q=\sqrt{6}$. Similarly, the left-hand circularly polarized Stokes pulse couple
the levels $|1,-1\rangle\Leftrightarrow |2,0\rangle$ and
$|1,+1\rangle\Leftrightarrow |2,+2\rangle $, and the Rabi frequencies are
given by
\begin{subequations}
\begin{eqnarray}
  \Omega_{S-}(1, -1; 2, 0)&=& {\rm const}_2\times {\cal E}_{S-}^* \,,\\
  \Omega_{S-}(1, +1; 2, +2)&=& q \times{\rm const}_2\times {\cal E}_{S-}^* \,.
\end{eqnarray}
\end{subequations}

In summary, the Rabi frequencies can be expressed as
\begin{subequations}
\begin{eqnarray}
  \Omega_{p+}(0, 0; 1, -1)&=& A \,\exp{i\varphi_A}\,, \\
  \Omega_{p-}(0, 0; 1, +1)&=& B \,\exp{i\varphi_B}\,, \\
  \Omega_{S+}(1, -1; 2, -2)&=& q\, C \,\exp{i\varphi_C}\,, \\
  \Omega_{S+}(1, +1; 2, 0)&=& C \,\exp{i\varphi_C}\,, \\
  \Omega_{S-}(1, -1; 2, 0)&=& D \,\exp{i\varphi_D}\,, \\
  \Omega_{S-}(1, +1; 2, +2)&=& q \, D \,\exp{i\varphi_D}\,.
\end{eqnarray}
\end{subequations}
The phases $\varphi_X$ are constants, and the envelope functions $X$ are
non-negative. The laser pulses may be detuned from resonance by a frequency
$\Delta$, but we require that they are at multi-photon resonance.  The
interaction Hamiltonian in the basis $\big\{ |0, 0\rangle$, $|1, -1\rangle$,
$|1, +1\rangle$, $|2, -2\rangle$, $|2, 0\rangle$, $|2, +2\rangle \big\}$ reads
\begin{widetext}
\begin{equation}
\label{eq:H}
H={1\over 2}\hbar\left[
\begin{array}{cccccc}
0&A\exp i\varphi_A&B\exp i\varphi_B&0&0&0\\
A\exp -i\varphi_A&\Delta&0&qC\exp i\varphi_C&D\exp i\varphi_D&0\\
B\exp -i\varphi_B&0&\Delta&0&C\exp i\varphi_C&qD\exp i\varphi_D\\
0&qC\exp -i\varphi_C&0&0&0&0\\
0&D\exp -i\varphi_D&C\exp -i\varphi_C&0&0&0\\
0&0&qD\exp -i\varphi_D&0&0&0\\
\end{array}
\right]\,.
\end{equation}
\end{widetext}
The time-evolution of the system is governed by the Schr\" odinger equation
\begin{equation}\label{eq:Sch}
  i\hbar\frac{\partial}{\partial t}|\psi\rangle = H |\psi\rangle\,.
\end{equation}

In a STIRAP process, the Hilbert space is split to dark and bright subspaces
and the system evolves in the dark subspace.  These subspaces are formed of
the instantaneous eigenstates of the time-dependent Hamiltonian. The dark
states belong to the eigenvalue zero and they do not have projection to the
excited state(s) \cite{arimondo}. The bright states belong to nonzero
eigenvalues and they overlap with the excited state(s).  In the adiabatic
limit, the system evolves entirely in the dark subspace. Now we continue the
analysis of the Hamiltonian Eq.~(\ref{eq:H}) in this direction.  First, we
determine the dark and bright subspaces of the Hilbert space. By applying a
diagonal unitary transformation $U_1$ to the system, the number of independent
phases in the Hamiltonian can be reduced to one:
\begin{equation}
\label{eq:H1}
H_1 = U_1 H U_1^\dagger = {1\over 2}\hbar\left[
\begin{array}{cccccc}
0&A&Be^{i\varphi}&0&0&0\\
A&\Delta&0&qC&D&0\\
Be^{-i\varphi}&0&\Delta&0&C&qD\\
0&qC&0&0&0&0\\
0&D&C&0&0&0\\
0&0&qD&0&0&0\\
\end{array}
\right],
\end{equation}
where 
\begin{eqnarray} \label{eq:U1def}
  U_1&=&\textrm{diag}\big[e^{-i(\varphi_A+\varphi_C)}, e^{-i\varphi_C},
  e^{i(\varphi_D-2\varphi_C)}, 1, \nonumber \\
  &&\qquad\,\, e^{i(\varphi_D-\varphi_C)}, e^{i(2\varphi_D-2\varphi_C)}\big]
  \,, 
\end{eqnarray}
and the remaining phase factor 
\begin{equation}\label{eq:phidef}
  \varphi=\varphi_B-\varphi_A+\varphi_C-\varphi_D \,.
\end{equation}
It will turn out that this phase factor is indeed relevant, since it affects
the probability amplitudes of the final state.

As a next step we are going to find a time-dependent unitary transformation
that defines a new basis, in which the dark and bright subspaces of the
Hilbert space are separated. By applying rotations and diagonal unitary
transformations in a well-chosen order we arrive at the following unitary
transformation:
\begin{equation}
U=U_4 U_3 U_2,
\label{eq:Udef}
\end{equation}
where
\begin{equation}
U_2=\left[\begin{array}{cccccc}
1&0&0&0&0&0\\
0&1&0&0&0&0\\
0&0&1&0&0&0\\[3pt]
0&0&0&-{y\over qx}&{yk\over x}&-{yk^2\over qx}\\[5pt]
0&0&0&{y^2k\over qx}&{1\over x}&{y^2k^3\over qx}\\[5pt]
0&0&0&{yk^2\over q}&0&-{y\over q}\\[5pt]
\end{array}\right]\,,
\end{equation}
and
\begin{equation}
U_3=\textrm{diag}[\exp i\xi,1,1,1,\exp i\zeta,1]\,,
\end{equation}
\begin{equation}
U_4=\left[\begin{array}{cccccc}
\cos\alpha&0&0&0&0&\sin\alpha\\
0&1&0&0&0&0\\
0&0&1&0&0&0\\
0&0&0&1&0&0\\
-\sin\beta\sin\alpha&0&0&0&\cos\beta&\sin\beta\cos\alpha\\
-\cos\beta\sin\alpha&0&0&0&-\sin\beta&\cos\beta\cos\alpha
\end{array}\right]\,,
\end{equation}
where we made use of the definitions
\begin{subequations}
\label{eq:params}
\begin{eqnarray}
  k&=&C/D\,, \label{eq:kdef} \\
  x&=&\sqrt{k^4+k^2q^2+1}/\sqrt{k^4+1}\,,\\
  y&=&-q/\sqrt{k^4+1}\,,\\
  \tan\alpha&=&{
    (k^4+1)yD
    \over \sqrt{A^2k^2+B^2-2ABk\cos\varphi}}\,,\\
  \tan\beta&=&-{y\over x}\cos\alpha\sqrt{{A^2+B^2k^6+2ABk^3
      \cos\varphi\over A^2k^2+B^2-2ABk\cos\varphi}}\,,
  \label{eq:params:beta}
  \\
  \xi&=&\arctan\left(\frac{B\sin\varphi}{Ak-B\cos\varphi}\right)\,,\\
  \zeta&=&\xi+\arctan\left(\frac{Bk^3\sin\varphi}{A+Bk^3\cos\varphi}\right)\,.
\end{eqnarray}
\end{subequations}

If we transform the Hamiltonian Eq.~(\ref{eq:H1}) with the unitary
transformation $U$ in Eq.~(\ref{eq:Udef}) we obtain 
\begin{equation} \label{eq:tH1}
  \tilde{H}_1 = UH_1U^{\dag}\,,
\end{equation}
which has two rows and two columns with zero elements. This means that the
unitary transformation $U$ defines two dark states. The first one is given by
the fourth row of $U_2$
\begin{equation}\label{eq:D1}
  |D_1\rangle = {1\over\sqrt{C^4+D^4+q^2C^2D^2}}\left[
    \begin{array}{c}
      0 \\ 0 \\ 0 \\ D^2 \\ -qCD \\ C^2
    \end{array}
  \right]\,.
\end{equation}
The one-dimensional subspace corresponding to that dark state is not altered
by the other unitary transformations $U_3$ and $U_4$, and it is a
superposition of the levels with $J\smash{=} 2$. The sixth row of the matrix $U$
defines the second dark state
\begin{equation}\label{eq:D2}
  |D_2\rangle = \left[\begin{array}{c}
      -\cos\beta\sin\alpha\exp i\xi\\
      0\\
      0\\
      -{y^2k\over qx}\sin\beta\exp i\zeta +{yk^2\over
        q}\cos\beta\cos\alpha\\[5pt]
      -{1\over x}\sin\beta\exp i\zeta\\[5pt]
      -{y^2k^3\over qx}\sin\beta\exp i\zeta - {y\over q} \cos\beta\cos\alpha
    \end{array}
  \right]\,.
\end{equation}
The states $|D_1\rangle$ and $|D_2\rangle$ are indeed dark states, since they
do not have components among the $J\smash{=}1$ levels. 

To complete the study of the dark and bright subspaces of our system we should
diagonalize completely the Hamiltonian $\tilde{H}_1$ in Eq.~(\ref{eq:tH1}).
This leads to rather involved formulae, however, for our purposes it is enough
to show that the eigenvalues of the $4\times4$ nonzero block of $\tilde{H}_1$
are nonzero and the corresponding eigenstates are bright states. We have
verified numerically using Maple \cite{maple} that the previous assumptions
are true. We conclude that the states given by Eqs.~(\ref{eq:D1}) and
(\ref{eq:D2}) are the two dark states of our system and it has four bright
states.  The dark subspace is degenerate, so particular attention should be
payed when one studies the time-evolution of the system \cite{una1,kisb,kisa}.
In general, for time-dependent dark states there appear a nonadiabatic
coupling term in the transformed Hamiltonian which couple the dark states.  We
will return to this issue in the Sec.~\ref{poptrans}. We can transform the
Schr\" odinger equation (\ref{eq:Sch}) to the time-dependent basis defined by
$U_1$ and $U$ in Eqs.~(\ref{eq:U1def}) and (\ref{eq:Udef})
\begin{equation}\label{eq:Schad}
  i\hbar\frac{\partial}{\partial t}|\tilde{\psi}\rangle = \tilde{H}
  |\tilde{\psi}\rangle\,,
\end{equation}
through the relations
\begin{eqnarray}\label{schad}
  |\tilde{\psi}\rangle & = & U U_1 |\psi\rangle\,, \nonumber \\
  \tilde{H} & = & \tilde{H}_1  + i \dot{U}U^{\dag}\,,
\end{eqnarray}
where $\tilde{H}_1$ is given by Eq.~(\ref{eq:tH1}). In this form the dark and
bright subspaces are separated. Since we want to design a STIRAP-like
population transfer process, in the following we focus our attention to the
dark states.  The bright states are needed when one wants to study the
coupling between the dark and bright subspaces, but since the formulae for the
bright states are too complicated, we shall do it in an other way.

\section{Adiabatic population transfer} \label{poptrans}

Now we make a further assumption: We assume that the ratios $A/B$ and $C/D$
are constant during the time-evolution, i.e. the envelope function of the two
pump pulses are the same, only their amplitudes and phases are different.
Similarly for the Stokes pulses. Consequently, the parameter $k$ defined in
Eq.~(\ref{eq:kdef}) is constant. Therefore, the dark state $|D_1\rangle$ is
constant throughout the whole time. This simplifies considerably the further
analysis. It follows immediately, that though the dark subspace is degenerate,
the two dark states do not mix with each-other during the interaction, since
$\langle D_2|\dot{D}_1\rangle\!=\!0$. Therefore, the Hamiltonian $\tilde{H}$
has a $2\times2$ zero matrix block acting on the dark subspace.

The dark state $|D_2\rangle$ has components among the $J\smash{=}0$ and
$J\smash{=}2$ states as well.  Hence, it is a good candidate to realize the
transfer process under consideration.  Initially the system is in the state
$|0, 0\rangle$. For implementing a STIRAP process, that state should be a dark
state of the system. Since the state $|D_1\rangle$ cannot participate in a
superposition forming the initial state, we require $|D_2\rangle$ to coincide
with the state $|0, 0\rangle$ at early times.  This condition is
satisfied if $C^2+D^2 \gg A^2+B^2$, which means that the pulses $C$ and $D$
arrive before $A$ and $B$. Though we have not presented explicitly the bright
states of our system, it is sure that they have no projection to the initial
state, since they are always orthogonal to the dark states and
$|D_2\rangle\equiv|0, 0\rangle$ initially, if the above condition is
fulfilled. This argument ensures that the initial state of the system lays in
the dark subspace which is necessary for a STIRAP process.

We have required that the final state of the system is a linear combination of
the levels with $J\smash{=} 2$ and $m\smash{=} -2, 0, +2$. The dark state
$|D_2\rangle$ has these components as well, so the previous condition is true
only if the first element of the state $|D_2\rangle$ is zero at the end of the
adiabatic evolution.  Consequently, at late times it is eligible that $C^2+D^2
\ll A^2+B^2$, therefore, the pulses $C$ and $D$ terminate before $A$ and $B$.

The previous conditions suggest that for successful population transfer we
need counterintuitive time ordering of the pump and Stokes pulses, similarly
to the original STIRAP. Now we consider the possibility of adiabatic
population transfer using the connectivity argument of Ref.~\cite{Martin1}:
The system Hamiltonian has a constant zero eigenvalue which is degenerate.
However, as we have shown in the beginning of this Section, the nonadiabatic
coupling between the two degenerate dark states vanishes because one of them
is constant. As we have seen in the previous paragraphs, the time dependent
dark state coincides with the initial state of the system at the beginning of
the process when only the Stokes pulse is present. Similarly, this dark state
lays in the required final state subspace at the end of the process when only
the pump pulse is present. Therefore, the connectivity requirements are
satisfied \cite{Martin1}, we have got a dark state that connects smoothly the
initial state of the system with the final state. However, besides these
conditions we have to fulfill adiabaticity as well. We will discuss this
issue below.

The considered six-level system can be regarded as four coupled three-level
systems: For $A=0$ and $C=0$ the pulses $B$ and $D$ define a three-level
STIRAP process between the levels $|0, 0\rangle$, $|1, +1\rangle$ and $|2,
0\rangle$. For $A=0$ and $D=0$ the pulses $B$ and $C$ define a three-level
STIRAP process between the levels $|0, 0\rangle$, $|1, +1\rangle$ and $|2,
+2\rangle$.  For $B=0$, $C=0$ and $B=0$, $D=0$ the situation is similar. The
envelopes $A$, $B$, $C$, $D$ can be parametrized in a form that emphasizes the
presented qualitative picture:
\begin{subequations}
\begin{eqnarray}
A&=&R_p\cos\eta\,,\\
B&=&R_p\sin\eta\,,\\
C&=&{R_S\over \sqrt{1+q^2}}\cos\nu\,,\\
D&=&{R_S\over \sqrt{1+q^2}}\sin\nu\,,
\end{eqnarray}
\end{subequations}
where the constant angles $\eta$ and $\nu$ are in the interval $[0,\pi/2]$.
These angles define the polarization of the laser pulses.  The six-level
system simplifies to a three-level $\Lambda$ system when $\eta,\nu = 0,
\pi/2$. For arbitrary constant angles $\eta$ and $\nu$ fixed couplings exist
between the individual three-level STIRAP processes.

If the time-evolution is adiabatic, then the dark and bright subspaces do not
mix during the whole interaction time \cite{una1,kisb,kisa}. Therefore, if the
initial state of the system lays in the dark subspace, it will remain there.
The physical consequence of the adiabatic evolution is that the excited states
$|1, \pm 1\rangle$ are only minimally populated throughout the transfer
process. As we saw in the previous paragraph, when $\eta, \nu = 0, \pi/2$ the
system reduces to the ordinary three-level STIRAP. In this case the
adiabaticity constraints are simple: Let us take $\eta, \nu = 0, \pi/2$, then
adiabaticity requires
\begin{equation}\label{adia1}
  \left|\dot{\vartheta}\frac{\sin^2\varrho}{\cos\varrho}\right|\ll 
  \frac{1}{2} \Omega\,,\qquad
  \left|\dot{\vartheta}\frac{\cos^2\varrho}{\sin\varrho}\right|\ll 
  \frac{1}{2} \Omega\,,
\end{equation}
where $\Omega = \sqrt{\Omega_p^2+\kappa^2\, \Omega_S^2}$, $\tan\vartheta =
\Omega_p/\kappa \Omega_S$, ($\kappa = 1, q$), and $\tan2\varrho =
\Omega/\Delta$. These conditions say that we need large pulse amplitudes and
smooth, long pulses. For a complete study we would need the explicit form of
the bright states $|B_k\rangle, k=1\hdots4$ of the Hamiltonian in
Eq.~(\ref{eq:H1}).  Then, adiabaticity requires to satisfy the following four
equations:
\begin{equation}\label{adia2}
  |\langle D_2|\dot{B}_k\rangle|\ll|\varepsilon_k - \varepsilon_0|\,,
  \qquad k=1\hdots4\,,
\end{equation}
where $\varepsilon_k$ is the adiabatic eigenenergy associated with
$|B_k\rangle$, and $\varepsilon_0=0$ belongs to the dark state $|D_2\rangle$.
These equations yield the {\em necessary} and {\em sufficient} conditions for
the pulses to be fulfilled in order that the evolution satisfy adiabaticity.
Although the bright states $|B_k\rangle$ are not available analytically, these
equations can be used to test adiabaticity numerically for a certain choice of
the pulses. In the numerical simulations presented below we used these
equations to test adiabaticity. Besides these conditions, below we show that a
simplified approach is also possible: we formulate a {\em sufficient}
condition for adiabatic evolution in the six-level system.

Our physical intuition suggests that if the three-level STIRAP processes for
the special polarizations $\eta,\nu = 0, \pi/2$~ are adiabatic, then the
six-level STIRAP process will also be adiabatic for constant angles $\eta$ and
$\nu$.  This is a {\em sufficient} condition for adiabaticity. It is
independent of the polarization of the pulses, in fact, we apply the
adiabaticity condition of the original three-level STIRAP to our six-level
scheme. We have performed numerical simulations to reveal whether this
assumption is valid or not.

\begin{figure}
\includegraphics[angle=-90,width=\textwidth/2-1cm]{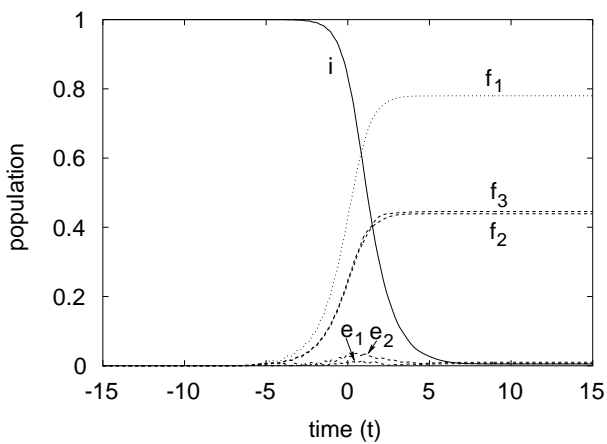}
\caption{The numerically computed time evolution of the populations are
  shown for a set of Gaussian pulses. The pulse shapes are the following:
  $R_p=10\exp[-((t-1.6)/2.8)^2]$, $R_S=10\sqrt{1+q^2}\exp[-((t+1.6)/2.8)^2]$;
  the polarizations are characterized by the angles $\eta=0.5$, $\nu=1.04$;
  finally $\varphi=3.34$. The excitation is resonant $\Delta=0$. Time is
  measured in arbitrary units.}
\label{fig:timeevol1}
\end{figure}

\begin{figure}
\includegraphics[angle=-90,width=\textwidth/2-1cm]{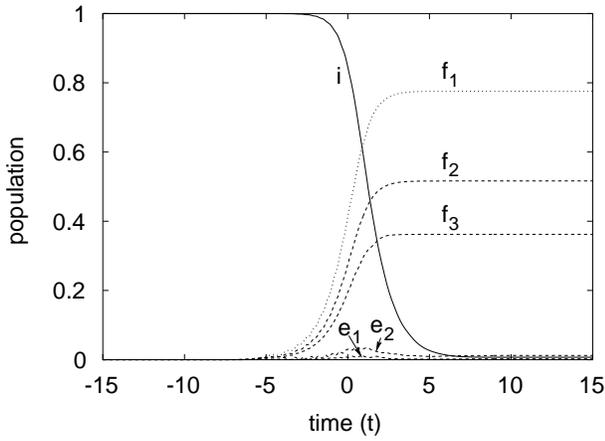}
\caption{The numerically computed time evolution of the populations are
  shown for another set of Gaussian pulses. The pulse shapes are the same as
  in Fig. \ref{fig:timeevol1}, only the phase $\varphi$ is changed to $4.74$.}
\label{fig:timeevol2}
\end{figure}

In Figs. \ref{fig:timeevol1} and \ref{fig:timeevol2} 
the numerically calculated time evolution of the populations are shown for
two different parameter sets. The pulses are Gaussian. The two parameter sets
differ only in the phase $\varphi$. The final populations, denoted by $f_1$,
$f_2$ and $f_3$ in the figures, are different, hence the phase $\varphi$ has
indeed impact on the magnitude of the probability amplitudes in the final
state. We interpret this phase dependence as a manifestation of quantum
interference among the rivaling coupled population transfer processes.

\begin{figure}
\includegraphics[angle=-90,width=\textwidth/2-1cm]{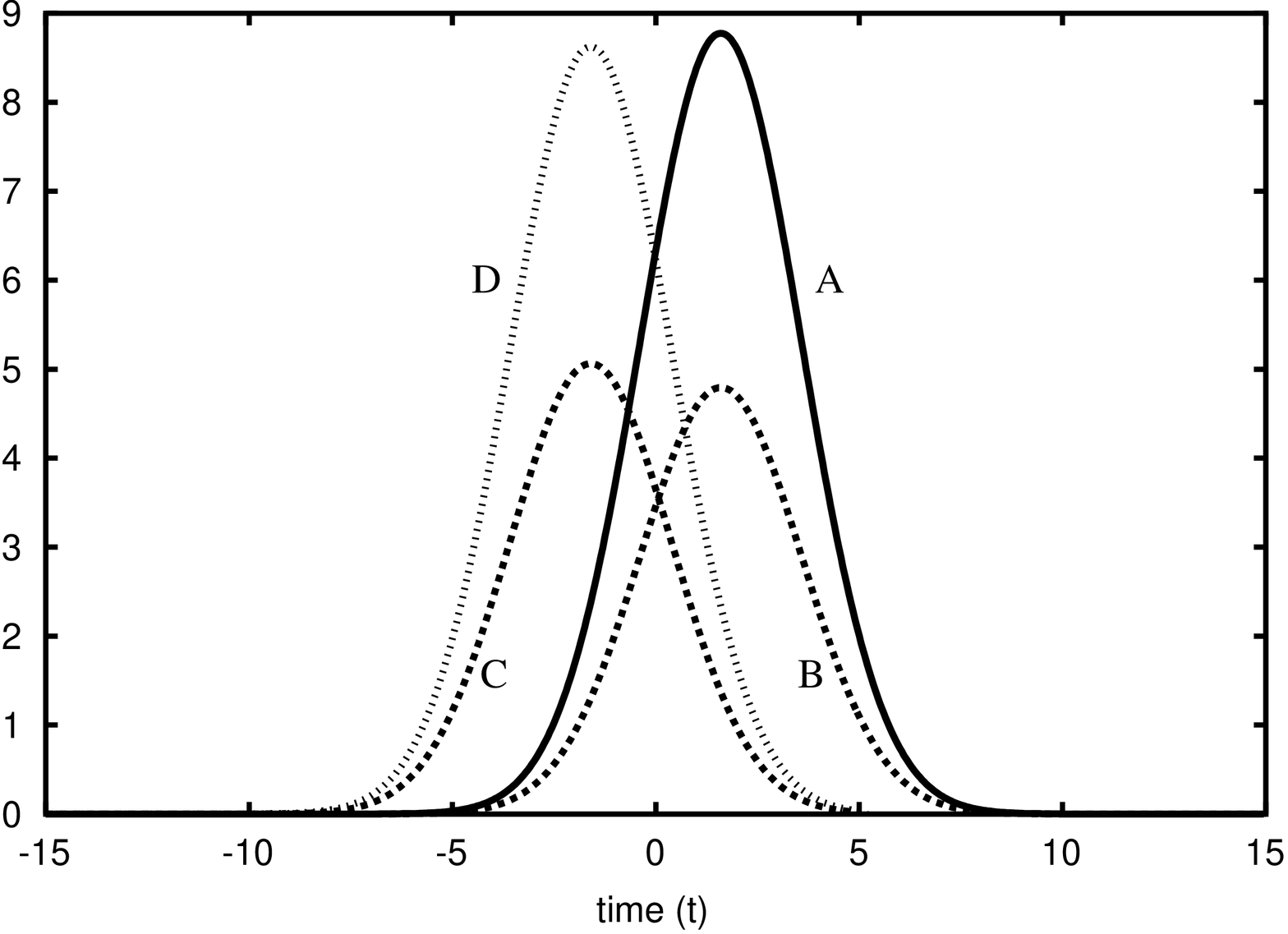}\\
\includegraphics[angle=-90,width=\textwidth/2-1cm]{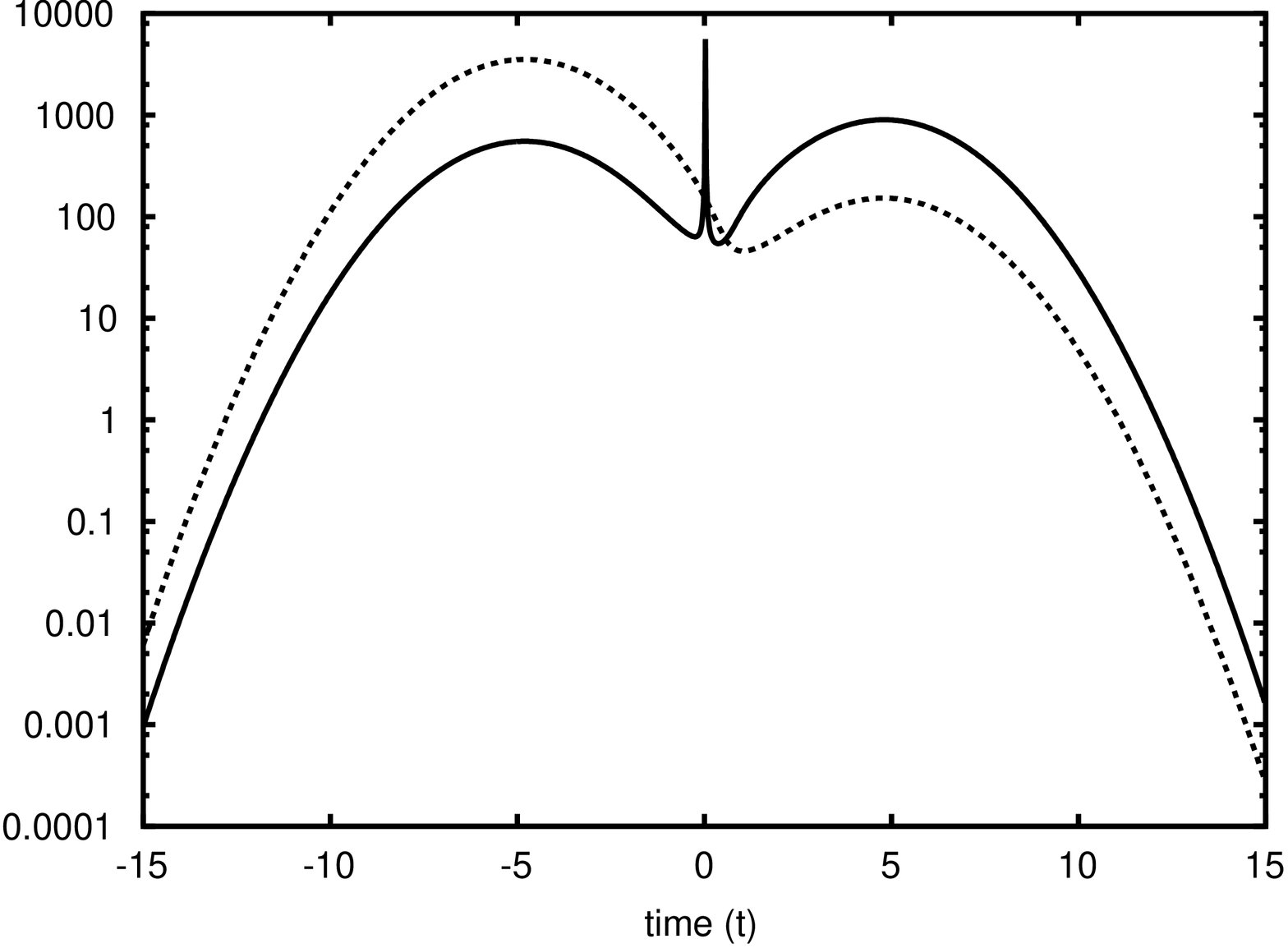}
\caption{The adiabaticity of the process shown in
  Fig.~\ref{fig:timeevol1}. The upper plot shows the pulses, in the lower
  figure the ratios $|\varepsilon_k-\varepsilon_0|/|\langle
  D_2|\dot{B}_k\rangle|\,,\, k=1\hdots4$ are plotted. Note that two pairs of
  the eigenstates and matrix elements coincide, hence only two curves can be
  distinguished. Time and frequency are measured in arbitrary units.}
\label{fig:adia1}
\end{figure}

To test adiabaticity in the previous examples, we evaluated Eqs.~(\ref{adia2})
for the parameters of Fig.~\ref{fig:timeevol1}. The results are shown in
Fig.~\ref{fig:adia1}. It can be seen that for that time-interval when
adiabaticity is expected the conditions are well fulfilled. The
non-adiabaticity of our six-level STIRAP process can be also quantified by the
maximum population of the bright subspace during time-evolution. This
non-adiabaticity is $0.014$ and $0.011$ for the two presented processes.  The
small values show that the mixing between the dark and bright subspaces is
negligible, therefore, the time-evolution of the system is adiabatic.  The
efficiency of the population transfer can be defined as the sum of the norm of
the final states with $J=2$.  For the two presented processes it is found to
be $99.98\%$ and $99.97\%$, showing that it is possible to perform a STIRAP
process in the considered six-level system.

The final state of the system in the adiabatic limit can be obtained
analytically from the expressions of the unitary transformations $U$ and
$U_1$ at the beginning and at the end of the population transfer
\begin{equation}\label{eq:psif}
  |\psi_f\rangle=U_1^\dagger U^\dagger_f U_i U_1| \psi_i\rangle\,,
\end{equation}
provided that $| \psi_i\rangle$ lays in the dark subspace.  In our case the
initial state of the system is $|\psi_i\rangle=|0, 0\rangle$.  After some
calculation we obtain
\begin{equation}\label{eq:psife}
  |\psi_f\rangle=\left[\begin{array}{c}
      0\\
      0\\
      0\\
      (-{y^2k\over qx}\sin\beta e^{-i\zeta}+{yk^2\over q}\cos\beta)
      e^{-i(\varphi_A+\varphi_C-\xi)} 
      \\
      {1\over x}\sin\beta e^{-i(\varphi_A+\varphi_D-\xi)} 
      \\
      (-{y^2k^3\over qx}\sin\beta e^{-i\zeta}-{y\over q}\cos\beta)
      e^{-i(\varphi_A-\varphi_C+2\varphi_D-\xi)} 
    \end{array}\right],
\end{equation}
where the parameters $k$, $x$, $y$, $\beta$, $\zeta$, $\xi$ are defined by
Eqs.  (\ref{eq:params}), and the parameters $\beta$, $\zeta$, and $\xi$ are
evaluated at the final time. The equations (\ref{eq:psif}) and
(\ref{eq:psife}) together with (\ref{eq:params}) show that our population
transfer process is robust with respect to small fluctuations of the
experimental parameters such as pulse shapes and pulse areas. However, the
relative phases of the pulses influence not only the phases in the final
superposition state but the probability amplitudes as well.

It is necessary to verify that all possible linear combinations of the states
with $J \smash{=} 2$ and $m\smash{=} -2, 0, +2$ can be created by the considered STIRAP
process.  Suppose, we have a prescribed final state
\begin{equation}
  |\psi_c\rangle=\left[\begin{array}{c}
      0\\
      0\\
      0\\
      c_1 e^{i\varphi_1}\\
      c_2 e^{i\varphi_2}\\
      c_3 e^{i\varphi_3}\\
    \end{array}\right],
\end{equation}
and we have found a pulse set with $\varphi_A=0$, $\varphi_B=\varphi$,
$\varphi_C=0$, $\varphi_D=0$ for which the final populations of the state
$|\psi_f\rangle$ are the same as in the prescribed state, and the equation
\begin{eqnarray}
\varphi_1+\varphi_3 - 2\varphi_2 &\equiv& \arg |\psi_f\rangle_4+\arg
|\psi_f\rangle_6-\nonumber\\
&&-2\arg |\psi_f\rangle_5 \mod 2\pi
\end{eqnarray}
holds. By adjusting the phases of the pulses to
\begin{subequations}
\label{eq:trans}
\begin{eqnarray}
\varphi_A&=&\arg|\psi_f\rangle_5-\varphi_2,\\
\varphi_B&=&\varphi+\arg|\psi_f\rangle_6-\varphi_3,\\
\varphi_C&=&\arg|\psi_f\rangle_5-\arg|\psi_f\rangle_6+\varphi_3-\varphi_2,\\
\varphi_D&=&0,
\end{eqnarray}
\end{subequations}
the phases of the final state change to the phases of the prescribed state.
The phase $\varphi$ defined by Eq.~(\ref{eq:phidef}) is invariant under the
presented phase adjustment, so the final populations remain the same.  It
follows that for deciding whether all prescribed final states can be reached,
it is enough to show that all possible amplitudes $(c_1, c_2, c_3)$ and all
possible values of
\begin{equation}
  \delta=\varphi_1+\varphi_3-2\varphi_2\mod 2\pi,
\end{equation}
can be reached by pulse sets with $\varphi_B=\varphi$ and
$\varphi_A=\varphi_C=\varphi_D=0$.

The amplitudes $(c_1, c_2, c_3)$ satisfy $c_1^2+c_2^2+c_3^2=1$. This triplet
defines a point on the one-eights of the surface of the unit sphere, since the
restriction $c_1,c_2,c_3\geq0$ must be satisfied. The surface points can be
parametrized by two polar angles $\theta\in[0,\pi/2]$ and $\chi\in[0,\pi/2]$,
therefore we have
\begin{subequations}
\begin{eqnarray}
c_1&=&\cos\theta \,,\\
c_2&=&\sin\theta\cos\chi \,,\\
c_3&=&\sin\theta\sin\chi \,. 
\end{eqnarray}
\end{subequations}

If we fix the shape of the pulses and the time-delay between them, then we
have three independent parameters which can be adjusted freely: the
polarizations $\eta$ and $\nu$ and the combination $\varphi$ of their relative
phases.  The final state can be characterized by the triplet $(\theta, \chi,
\delta)$.  Mathematically, our population transfer process assigns to each
triplet $(\eta,\nu,\varphi)$ a triplet $(\theta, \chi, \delta)$.  If these
points fill the cube $[0\ldots\pi/2,0\ldots\pi/2,0\ldots2\pi]$, then all
prescribed states can be reached by choosing a corresponding triplet
$(\eta,\nu,\varphi)$ and performing the transformation defined in Eq.
(\ref{eq:trans}) on the initial phases of the pulses.

\begin{figure}[p]
\includegraphics[angle=-90,width=\textwidth/2-1cm]{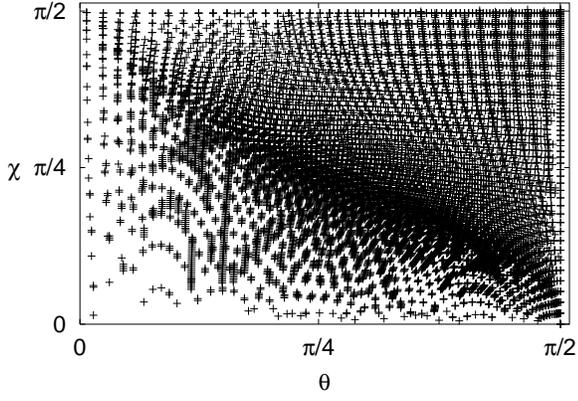}
\caption{The slice with $\delta=0.30\pm0.05$ of the cube filled by the
  points $(\theta,\chi,\delta)$ which result from the numerical solution of
  the Scr{\" o}dinger equation (\ref{eq:Sch}). We have fixed the shape of the
  pulses and the time delay between them, only their polarizations and
  relative phases were varied. }
\label{fig:num0.30}
\end{figure}

\begin{figure}[p]
\includegraphics[angle=-90,width=\textwidth/2-1cm]{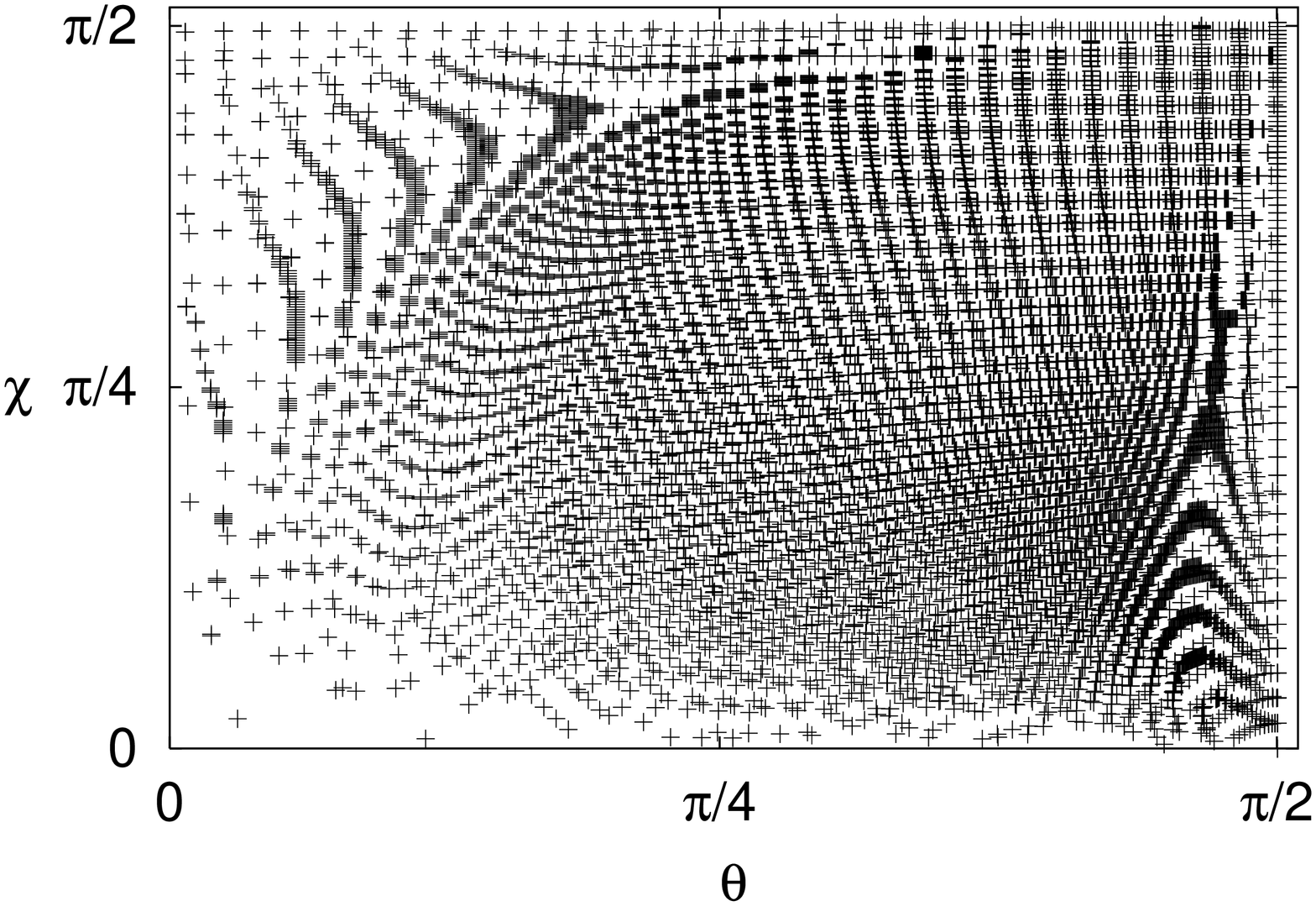}
\caption{Same as Fig.~\ref{fig:num0.30} but $\delta=2.4$. } 
\label{fig:num2.40}
\end{figure}

\begin{figure}[p]
\includegraphics[angle=-90,width=\textwidth/2-1cm]{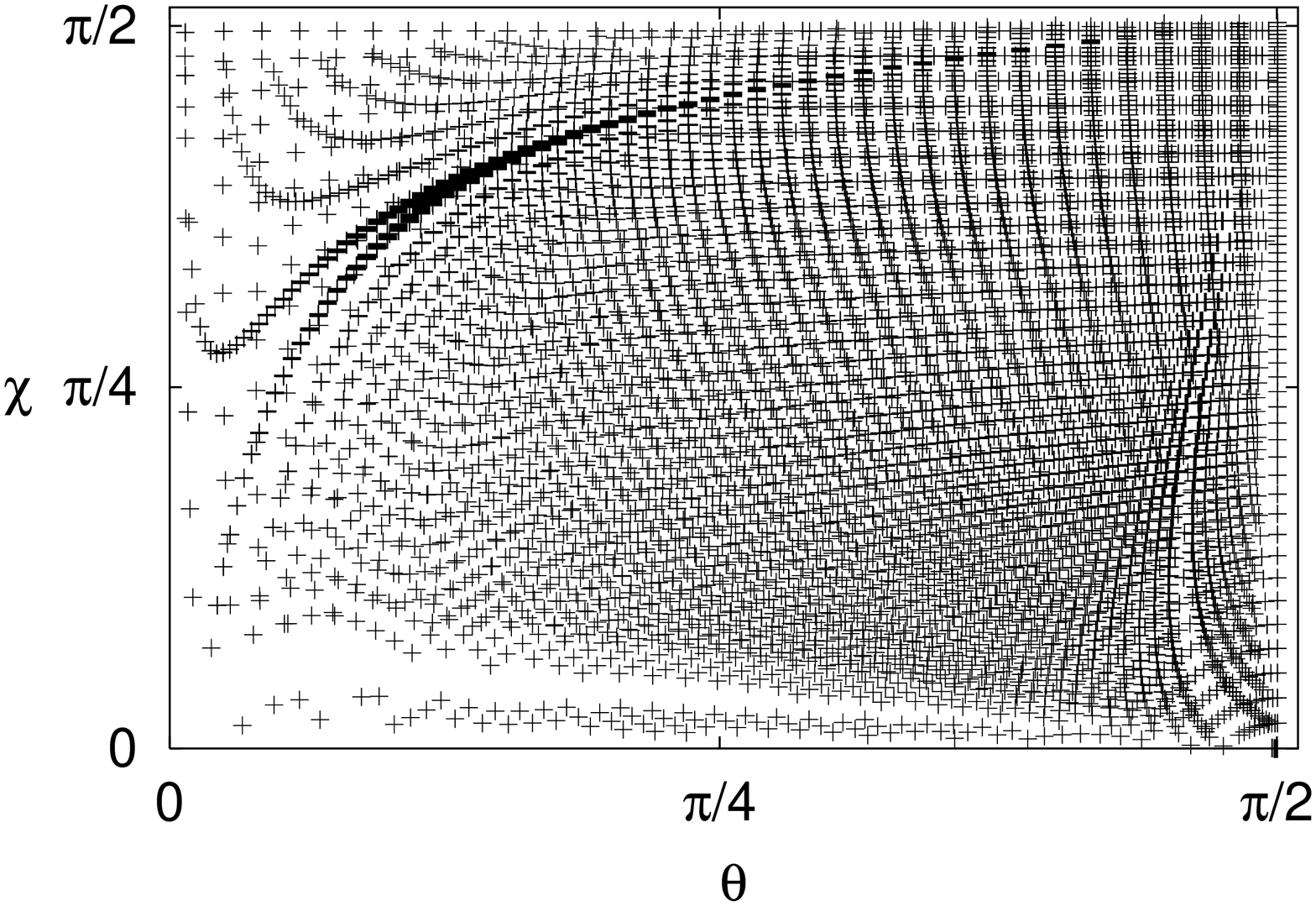}
\caption{Same as Fig.~\ref{fig:num0.30} but $\delta=3.2$. } 
\label{fig:num3.20}
\end{figure}

\begin{figure}[p]
\includegraphics[angle=-90,width=\textwidth/2-1cm]{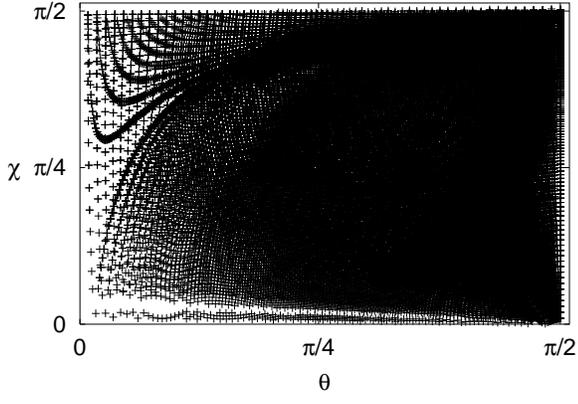}
\caption{Same as Fig.~\ref{fig:num3.20} but the points are computed from the
  analytical solution Eq.~(\ref{eq:psife}). } 
\label{fig:numa3.20}
\end{figure}

\begin{figure}[p]
\includegraphics[angle=-90,width=\textwidth/2-1cm]{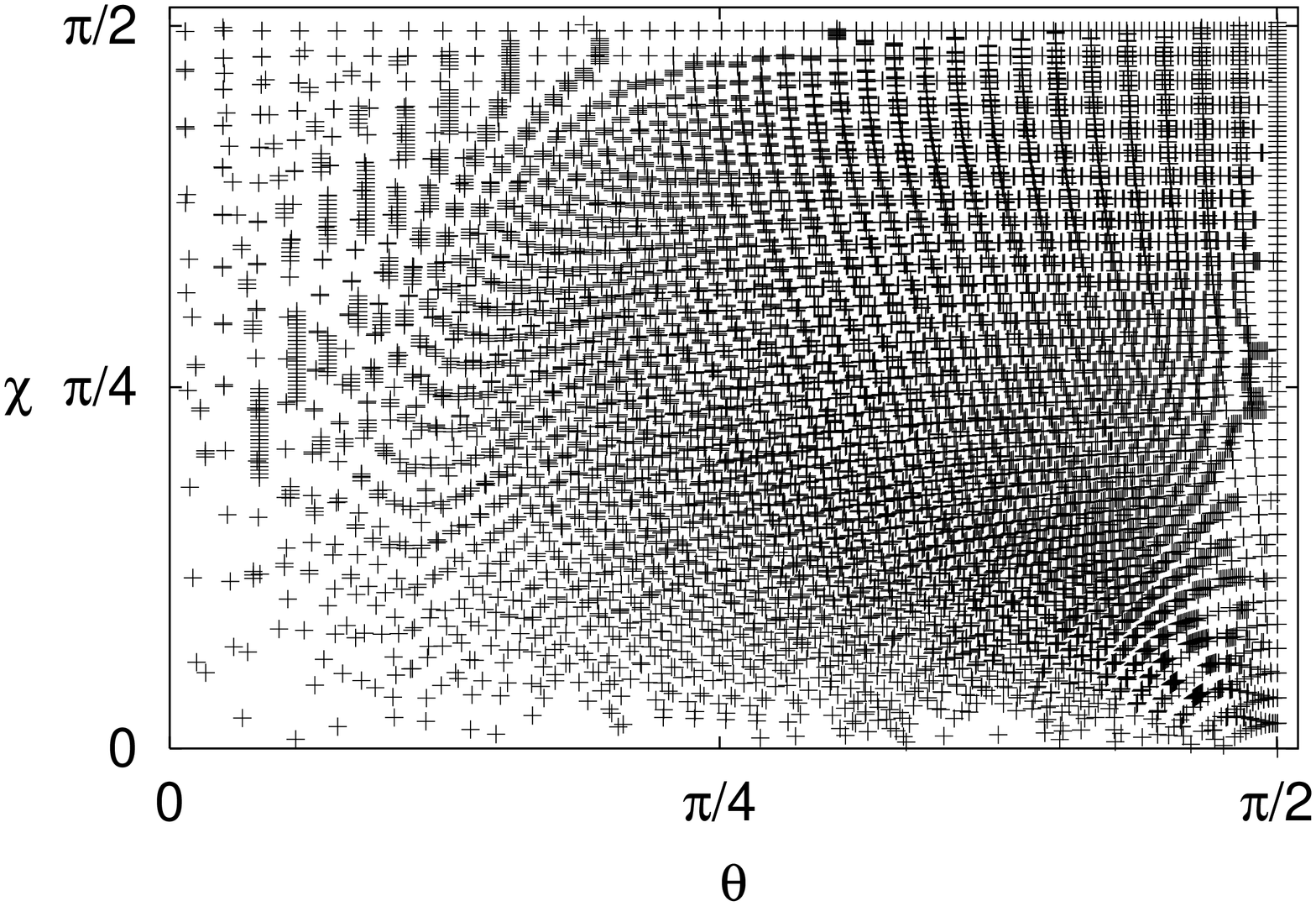}
\caption{Same as Fig.~\ref{fig:num0.30} but $\delta=4.5$. } 
\label{fig:num4.50}
\end{figure}

\begin{figure}[p]
\includegraphics[angle=-90,width=\textwidth/2-1cm]{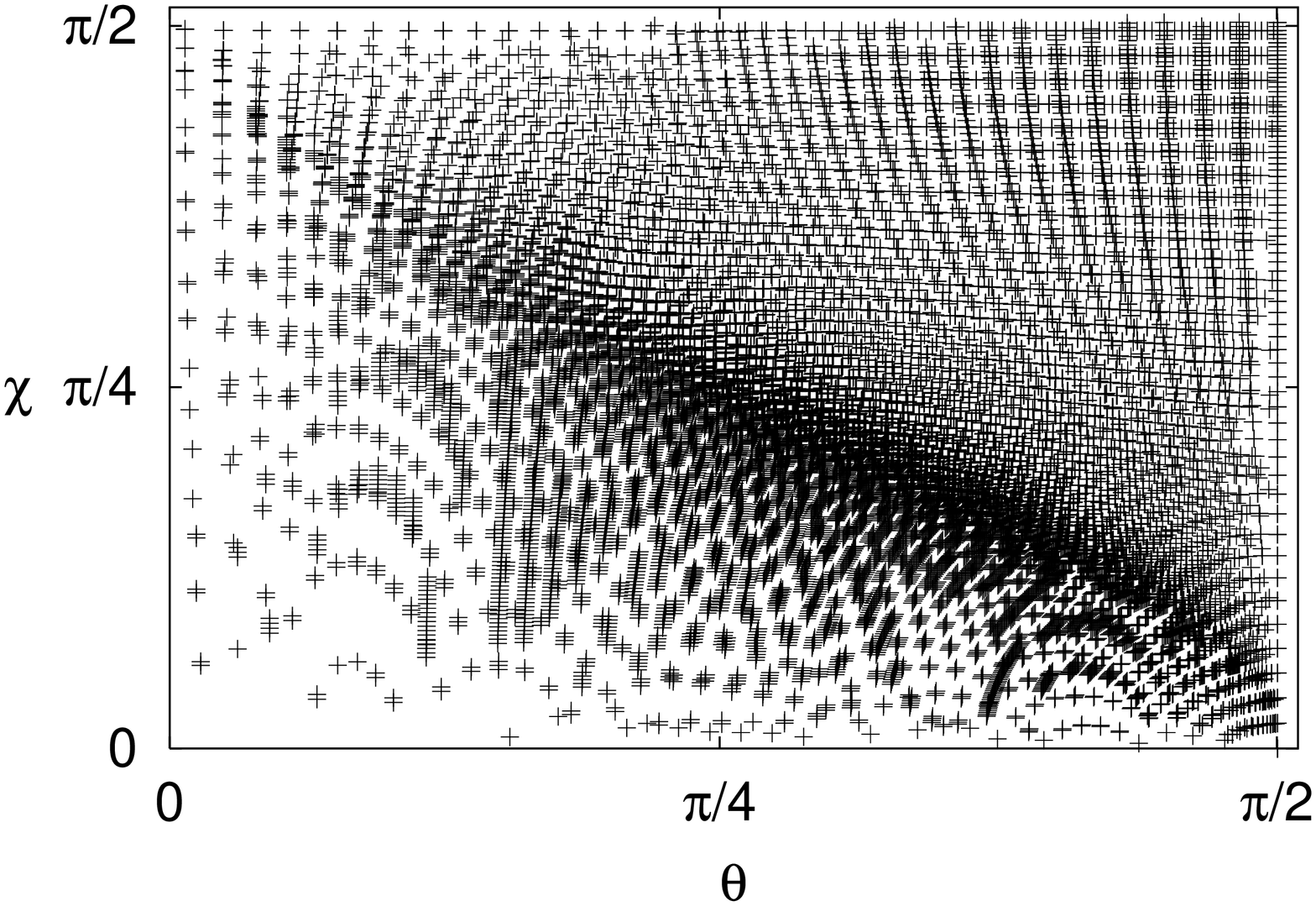}
\caption{Same as Fig.~\ref{fig:num0.30} but $\delta=6.0$. } 
\label{fig:num6.00}
\end{figure}

We have performed numerical simulations in order to answer the above question.
The angles $\eta$, $\nu$ and $\varphi$ were incremented by 0.022 in their
domain range. We have used the envelope functions
$R_p=15\exp[-((t-1.8)/2.82)^2]$ and $R_S=15\sqrt{1+q^2}\exp[-((t+1.8)/2.82)^2]$ for the
pump and Stokes pulses, respectively. We have verified that these pulses
fulfill the adiabaticity conditions Eq.~(\ref{adia1}) for the field
polarizations $\eta, \nu=0, \pi/2$. The resulting triplet set was divided into
slices of width 0.1 in $\delta$. Each slice was checked whether it is filled
with points. Some of the slices are shown in Figs.  \ref{fig:num0.30},
\ref{fig:num2.40}, \ref{fig:num3.20}, \ref{fig:numa3.20}, \ref{fig:num4.50},
\ref{fig:num6.00}.  We have found that each slice is equally and quite
uniformly filled with points.

We have compared the numerical results with the analytical solution given by
Eq.  (\ref{eq:psif}). Note that the analytical solution is valid in the
adiabatic limit. In this case the increments of the parameters $\eta$, $\nu$
and $\varphi$ were 0.015, giving a more detailed picture about the
distribution of the resulting points $(\theta,\chi,\delta)$. The differences
between the numerical simulation and analytical solution were small, below one
percent, showing that the chosen pulse-sets realized a nearly adiabatic
evolution. In Fig. \ref{fig:numa3.20} we show a slice of the distribution for
the same parameter set as in Fig. \ref{fig:num3.20}.

The maximum non-adiabaticity, that we measure as the maximum population of the
bright subspace, encountered during the simulation was 0.006. We have also
verified that the adiabaticity conditions Eq.~(\ref{adia2}) are fulfilled for
the whole parameter range. Therefore, our assumption is indeed true: if the
population transfer process is adiabatic for some special polarizations of the
pulses $\eta, \nu=0, \pi/2$, i.e. we have an effective three-level system,
then it remains adiabatic for arbitrary polarizations, i.e. all the six states
are coupled. This is a {\em sufficient} condition for the pulses to satisfy
adiabaticity. Based on the results of the numerical simulation and the
analytical solution, we conclude that all prescribed states can be reached by
the presented STIRAP-like population transfer process and the evolution is
adiabatic.

\section{Summary}\label{sum}
In this paper we have discussed a STIRAP-like population transfer process in a
six-level $\Lambda$ system. The initial state has angular momentum $J=0$, the
excited state has $J=1$ and the final state has $J=2$. Our aim is to create
coherent superposition states on the magnetic sublevels of the final state.
Similarly to the original STIRAP, the couplings are realized by a pump and a
Stokes laser pulse, which are elliptically polarized in our case. The shape of
the pulses and the time-delay between them are fixed.  However, their
polarizations and the relative phases are freely adjustable but also fixed
throughout the whole time. We have defined the Hamiltonian of this system in
the RWA approximation.  We have found that by time-independent unitary
transformations not all the phases can be eliminated from the Hamiltonian,
there remains one. This phase factor proved to be indeed relevant, since it
influences the probability amplitudes in the final superposition state. We
have determined the dark and bright subspaces of our system and we have given
explicitly the two dark states. We have shown that the desired population
transfer can be realized by a counterintuitive pulse sequence similarly to the
original STIRAP. We have studied the conditions for adiabatic evolution in our
system: In the case of special polarizations of the pulses the coupling
pattern reduces to a simple three-level system.  The adiabaticity conditions
are well known for this case.  We have shown numerically that the system
evolves adiabatically for general polarizations as well if adiabaticity
prevails for the special choice of the polarizations. We have compared the
results of the simulation with the analytically calculated final states in the
adiabatic limit, and an excellent agreement has been found. We have verified
numerically that the whole final state space can be covered by varying the
relative phases and the polarizations of the two exciting pulses, while the
pulse shapes and the time-delay between them are fixed.

As we mentioned before, the probability amplitudes in the final state depend
on the relative phases of the laser pulses. This is unusual in case of the
STIRAP process, up to our knowledge that is the first time when the relative
phases do matter. The phase dependence results from quantum interference,
because the coupling pattern can be considered as a coupled set of three-level
STIRAP systems. Therefore, the final state emerges from the rivaling STIRAP
processes.

Finally, we comment briefly the experimental realization of our scheme: It
seems easy to find an atomic system where the coupling pattern can be
attained. Two elliptically polarized pulses are required, however, the phase
shift between the left- and right-rotating components should be adjustable.
Practically this seems the only additional experimental effort to be done
compared with the realization of the traditional three-level STIRAP process.

\acknowledgments 

This work was supported by the European Union Research and Training Network
COCOMO, contract HPRN-CT-1999-00129.  Z.K. acknowledges the support of the
J\'anos Bolyai program of the Hungarian Academy of Sciences. A.K. acknowledges
the support of the Research Fund of Hungary under contract No. T034484.



\begin{thebibliography}{99}
%
\bibitem{LZ}{ L.D. Landau, Phys. Z. Sowjetunion {\bf 2}, 46 (1932); C. Zener,
  Proc. R. Soc. London Ser. A {\bf 137}, 696 (1932).}
\bibitem{Rosen}{ N. Rosen and C. Zener, Phys. Rev. {\bf 40}, 502 (1932).}
%
\bibitem{Demkov}{ Yu.N. Demkov and M. Kunike, Vestn
  Leningr. Univ. Fis. Khim. {\bf 16}, 39 (1969).} 
%
\bibitem{Hioe}{ F.T. Hioe, Phys. Rev. A {\bf 30}, 2100 (1984).}
%
\bibitem{Suominen}{ K.-A. Suominen, B.M. Garraway, and S. Stenholm,
  Opt. Commun. {\bf 82}, 260 (1991).}
%
\bibitem{vita}{ N.V. Vitanov, T. Halfmann, B.W. Shore, and K. Bergmann,
  Annu. Rev. Phys. Chem. {\bf 52}, 763 (2001).}
%
\bibitem{vitb}{ N.V. Vitanov, M. Fleischauer, B.W. Shore and K. Bergmann,
  Adv. At. Mol. Opt. Phys. {\bf 46}, 55 (2001).}
%
\bibitem{marte}{ P.~Marte, P.~Zoller and J.L.~Hall, Phys. Rev. A {\bf
    44}, R4118 (1991).}
%
\bibitem{una1}{ R.G. Unanyan, M. Fleischhauer, B.W. Shore, and K.
  Bergmann, Opt. Commun. {\bf 155}, 144 (1998).}
%
\bibitem{theuer}{ H. Theuer, R.G. Unanyan, C. Habscheid, K. Klein, and
  K. Bergmann, Opt. Express {\bf 4}, 77 (1999).}
%
\bibitem{una2}{ R.G. Unanyan, B.W. Shore, and K. Bergmann, Phys. Rev. A
  {\bf 59}, 2910 (1999).}

\bibitem{Chang} B.Y. Chang, I.R. Sol\'a, V.S. Malinovsky, and J. Santamar{\'
    \i}a, Phys. Rev. A {\bf 64}, 033420 (2001).
%
%
\bibitem{una3}{ R.G. Unanyan, B.W. Shore, and K. Bergmann, Phys. Rev. A
  {\bf 63}, 043401 (2001).}
%
\bibitem{kisc}{ Z. Kis and F. Renzoni, Phys. Rev. A {\bf 65}, 032318 (2002).}
%
\bibitem{kisd}{ Z. Kis and S. Stenholm, Phys. Rev. A {\bf 64}, 065401 (2001).}
%
%
\bibitem{kisa}{ Z. Kis and S. Stenholm, special issue of J. Mod.
  Optics {\bf 49}, 111 (2002).}
%
\bibitem{kisb}{ Z. Kis and S. Stenholm, Phys. Rev. A {\bf 64}, 063406 (2001).}
%
%
\bibitem{Shore} B.W. Shore, J. Martin, M.P. Fewell, and K. Bergmann,
  Phys. Rev. A {\bf 52}, 566 (1995).

\bibitem{Martin1} J. Martin, B.W. Shore, and K. Bergmann,
  Phys. Rev. A {\bf 52}, 583 (1995).

\bibitem{Martin2} J. Martin, B.W. Shore, and K. Bergmann,
  Phys. Rev. A {\bf 54}, 1556 (1996).

\bibitem{Carroll} C.E. Carroll and F.T. Hioe, Phys. Rev. Lett. {\bf 68}, 3523
  (1992). 

\bibitem{Nakajima} T. Nakajima, M. Elk, J. Zhang, and P. Lambropoulos,
  Phys. Rev. A {\bf 50}, R913 (1994).

\bibitem{vitc} N.V. Vitanov and S. Stenholm, Phys. Rev. A {\bf 60}, 3820 
  (1999). 
  
\bibitem{Yatsenko} L.P. Yatsenko, R.G. Unanyan, K. Bergmann, T. Halfmann, and
  B.W. Shore, Opt. Commun. {\bf 135}, 406 (1997).

\bibitem{vitd} N.V. Vitanov and S. Stenholm, Phys. Rev. A {\bf 56}, 741
  (1997). 

\bibitem{Paspalakis} E. Paspalakis, M. Protopapas, and P.L. Knight,
  Opt. Commun. {\bf 142}, 34 (1997).

\bibitem{una4} R.G. Unanyan, N.V. Vitanov, and S. Stenholm, Phys. Rev. A {\bf
    57}, 462 (1998). 
  
\bibitem{una5} R.G. Unanyan, N.V. Vitanov, B.W. Shore, and K. Bergmann, Phys.
  Rev. A {\bf 61}, 043408 (2000).

\bibitem{bruce} B.W. Shore, {\em The theory of coherent atomic excitation}
(Wiley, New York, 1990).

\bibitem{arimondo}{ E. Arimondo, in Progress in Optics ed. E. Wolf,
  XXXV, p. 257 (Elsevier, Amsterdam, 1996).}
%
%
\bibitem{maple}{ Maple is a symbolic computer algebra software, a product of
    Waterloo Maple Inc.}
%
%
\end{thebibliography}
\end{document}